 \documentclass[pmlr,twocolumn]{jmlr} 

\usepackage{caption}

\usepackage{makecell}

\usepackage{booktabs}
\usepackage[load-configurations=version-1]{siunitx} 

\theorembodyfont{\upshape}
\theoremheaderfont{\scshape}
\theorempostheader{:}
\theoremsep{\newline}

\jmlrvolume{ML4H Extended Abstract Arxiv Index}
\jmlryear{2020}
\jmlrsubmitted{2020}
\jmlrworkshop{Machine Learning for Health (ML4H) 2020} 

\title[GloFlow: Whole Slide Images for Pathology from Video]{GloFlow: Global Image Alignment for Creation of Whole Slide Images for Pathology from Video}

\makeatletter
\newcommand{\printfnsymbol}[1]{%
  \textsuperscript{\@fnsymbol{#1}}%
}
\makeatother

\author{%
\Name{Viswesh Krishna}\thanks{Equal Contribution} \Email{viswesh@stanford.edu}\\
\Name{Anirudh Joshi}\footnotemark[1] \Email{anirudhjoshi@stanford.edu}\\
\Name{Philip L. Bulterys} \Email{bulterys@stanford.edu}\\
\Name{Eric Yang} \Email{ericyang@stanford.edu}\\
\Name{Andrew Y. Ng} \Email{ang@cs.stanford.edu}\\
\Name{Pranav Rajpurkar} \Email{pranavsr@cs.stanford.edu}}

\begin{document}

\maketitle

\begin{abstract}
The application of deep learning to pathology assumes the existence of digital whole slide images of pathology slides. However, slide digitization is bottlenecked by the high cost of precise motor stages in slide scanners that are needed for position information used for slide stitching.
We propose GloFlow, a two-stage method for creating a whole slide image using optical flow-based image registration with global alignment using a computationally tractable graph-pruning approach.
In the first stage, we train an optical flow predictor to predict pairwise translations between successive video frames to approximate a stitch. In the second stage, this approximate stitch is used to create a neighborhood graph to produce a corrected stitch.
On a simulated dataset of video scans of WSIs, we find that our method outperforms known approaches to slide-stitching, and stitches WSIs resembling those produced by slide scanners.

\end{abstract}

\section{Introduction}
\label{sec:intro}

 \begin{figure*}[!ht]
   \includegraphics[width=\textwidth,height=6cm]{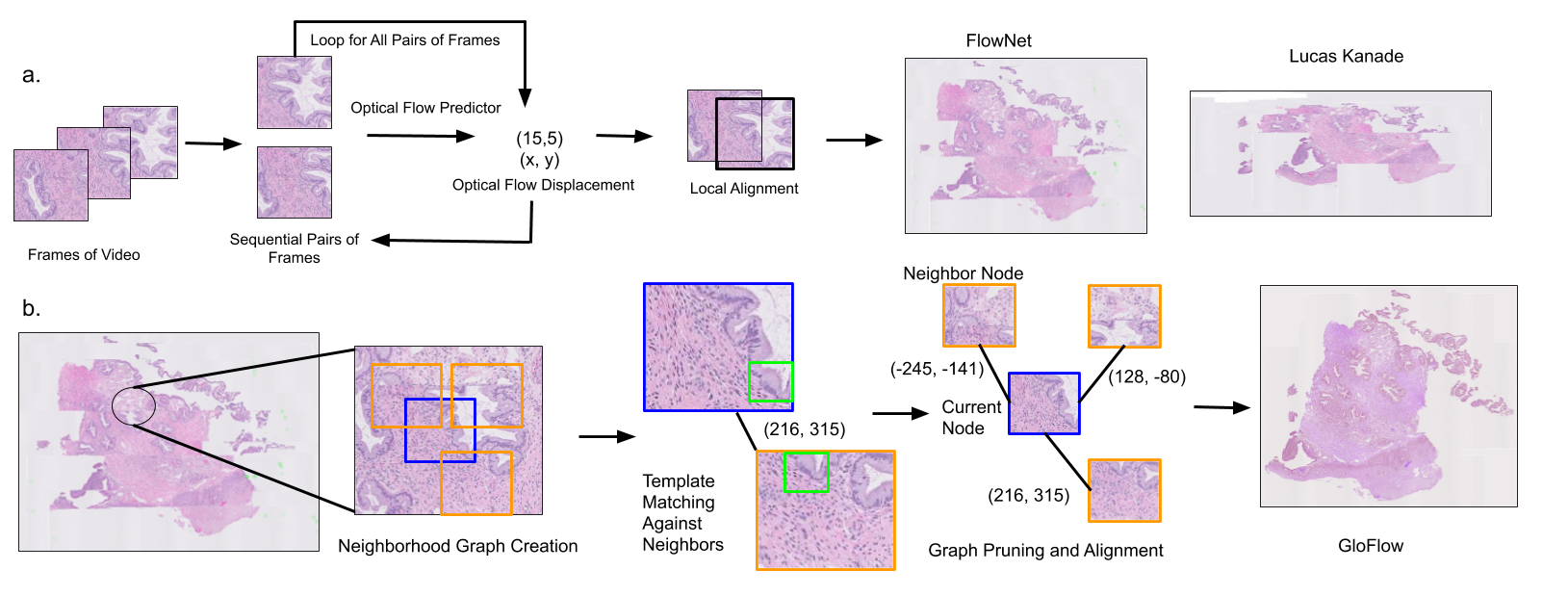}
  \caption{\textbf{Overview of GloFlow} (a) Optical flow is computed using pairwise sequential frames of video. Each pair of frames is aligned to produce an approximate stitch. (b) Using the approximate WSI, we create a neighborhood graph that uses template matching for refining the stitch.}
 \label{fig:main}
 \end{figure*}

Although the application of deep learning to pathology requires digital whole slide images, the majority of pathology slides across the world are not digitized \citep{hanna2019whole}. Conventional slide scanning technology uses precise sub-micron motor stages to capture image tiles at a magnification and stitch them together, and can cost upwards of \$70,000 \citep{isse2012digital, chalfoun2017mist, farahani2015whole}. An alternative is to use cheaper hardware with less precise position information and more sophisticated computer vision methods to stitch the slide together \citep{beckstead2003high, montalto2011autofocus, zarella2019practical}.

The creation of a whole slide image from a video scan can be cast as an image stitching problem. The image stitching problem in the context of producing mosaics and panoramas has been well studied: optical flow methods like Lucas Kanade and \citep{carozza2011incremental} FlowNet \citep{dosovitskiy2015flownet, ilg2017flownet} can perform pairwise image registration, as can homography based methods like RANSAC \citep{fischler1981random}. 
However, purely pairwise image registration can accumulate errors in the creation of the overall image stitch.
These errors can be corrected using global image alignment algorithms that optimize which pairwise registrations are more likely than others \citep{pellikka2020robust}. However these algorithms are computationally intractable as they attempt to produce pairwise translations between all possible images. \citep{yu2011automated}.

To tackle the slide-stitching problem, we propose a two-stage method for creating a whole slide image using optical flow-based image registration with global alignment using a computationally tractable graph-pruning approach.
In the first stage, we train an optical flow predictor to predict pairwise translations between successive video frames to approximate a stitch. In the second stage, this approximate stitch is used to create a neighborhood graph to produce a corrected stitch. GloFlow achieves an order of magnitude better Re-EPE (measure of global error) of 50.99 compared to Re-EPE of 745 of just pairwise optical flow. GloFlow is also twice as computationally efficient as other global alignment methods and produces whole slide images that closely resemble that taken from expensive scanners. 



Our proposed approach ensures that no position information is required during the stitching process and can be fast enough to deliver results within minutes. Our approach may enable the rapid digitization of whole slide images directly from widely-available microscopes.

\section{Methods}
The creation of WSIs from a video scan can be cast as an image stitching problem. The input of the task is a video over the slide where each frame of the video is a small section of the slide. The video is taken as described above and no information is provided as to when vertical row changes occur. The output for the task is a stitched WSI.

\subsection{Data Generation}
We develop a procedure to simulate video capture of slides under a microscope using digitized WSIs. This provides a simple testbed for the comparison of approaches and groundtruth for the whole-slide image stitch. We extract 512x512 patches of a digital WSI at constant magnification with small translational displacements between each other. In order to simulate the general movement of a slide under a microscope, we extract patches in a boustrophedonic manner - i.e. alternating left to right passes over the vertical length of the image beginning at the top-left corner. After each horizontal pass, we extract successive patches with vertical displacements till we cross a specified height after which we begin the next horizontal pass in the opposite direction. These images, when considered sequentially simulate a video capture of a slide moved under a microscope without associated local or global location data. Details of variation in data generation can be found in Appendix \ref{apd:first}.

\subsection{GloFlow}

 


We propose GloFlow, a two-stage method for creating a whole slide image using optical flow-based image registration with global alignment using a computationally tractable graph-pruning approach.
In the first stage, we train an optical flow predictor to predict pairwise translations between successive video frames to approximate a stitch. In the second stage, this approximate stitch is used to create a neighborhood graph to produce a corrected stitch.

\paragraph{Creating an Approximate Stitch Using Pairwise Displacements.} The task in the stage one is to form an approximate WSI. We compute pairwise translations using two optical flow methods. The first method is Lucas-Kanade (LK) optical flow \citep{lucas1981iterative, carozza2011incremental}, that uses Shi-Tomasi feature points. The second method is our modification of a deep learning-based flow predictor \citep{dosovitskiy2015flownet}. In this method called FlowNetMod (FNM), we train a neural network to output the translation in the $x$ and $y$ coordinates between two frames. FNM is our extension of FlowNet \citep{dosovitskiy2015flownet} that outputs a single flow prediction as opposed to a pixel-wise flow. This is because in our setting, we can assume that every pixel has the same displacement. The input to the network for each forward pass is two frames from the video concatenated channelwise and the $L1$ loss is computed using the predicted and groundtruth flows. During inference, a series of pairwise flow predictions between consecutive frames of a video are used to create the approximate WSI.

\paragraph{Correcting Approximate Stitch using Neighborhood Graph.}
In the second stage, the approximate stitch produced in the first stage is used to create a neighborhood graph that produces a corrected stitch. A neighborhood graph is created as follows. All frames within a radius of twice the frame width from the upper left coordinate of a chosen frame are considered to be neighbors of that frame. These neighbors are added as connected nodes to the original frame and this process is repeated for each frame in the approximate WSI to create a neighborhood graph.

Using this neigborhood graph, we redo image registration to produce a corrected stitch. For each frame, Shi-Tomasi feature point \citep{shi1994good} neighborhoods are aggregated via dilation to form templates which are matched with neighbors to create a set of possible translations from the frame for each neighbor. A multigraph is formed by considering the set of translations (forming directed edges) between a source frame and its neighbors.
A weight is assigned to each edge corresponding to the maximum correlation coefficient during template matching.
The multigraph is then pruned to form a directed graph by first aggregating similar translations and then removing edges with low weights.
In order to ensure that the translation between a pair of nodes is consistent, each pair of translations in the directed graph between a given pair of nodes are checked to be equal in magnitude and opposite in direction which guarantees that they correspond to the same translation and thus the directed graph is converted to an undirected graph.
Global coordinates are computed for every frame using the neighborhood translations to create final stitch.
In order to reduce computational costs but still ensure a high quality stitch, we sample every 20th frame from the original video.

\subsection{Comparisons and Evaluation Metrics}
We compare our two-stage GloFlow approach against stage one only and stage two only methods. Stage one only methods include
Lucas-Kanade (LK) optical flow, and FlowNetMod (FNM), described in a previous section. The Stage two only method uses graph-based alignment (Pure Graph) without an approximate stitch initialization.

Global coordinates for each frame, generated from the optical flow and graph alignment translations are used for error computation.
We introduce and report the Re-centered endpoint error (Re-EPE), a measure of the accuracy of stitching complete sections in the slide. Re-EPE extends the End Point Error (EPE), a standard error measure for optical flow evaluation; however, unlike EPE, Re-EPE is invariant to global offset between stitches. Re-EPE is the average of EPEs computed by iterating over nodes in a graph and recentering the predicted and groundtruth coordinates to that node. Equations are in Appendix \ref{apd:second}.


\section{Results}

We find that GloFlow significantly outperforms both optical flow methods (LK, FNM) and graph alignment (Pure Graph), as seen in Table \ref{tab:table1}. GloFlow with both LK and FNM as stage one have comparable Re-EPEs (GloFlow [LK] = 50.99, GloFlow [FNM] = 51.14) and are an order of magnitude more accurate than optical flow (LK = 1386.08, FNM = 745.80) and graph alignment (Pure Graph = 507.09) techniques used individually. We find that FNM achieves lower error than LK on both EPE (FNM = 6.92, LK = 12.52) and Re-EPE (FNM = 745.80, LK = 1386.08). FNM also produces a better approximate stitch as can be seen in Figure \ref{fig:main}. Measuring runtime, GloFlow takes 684.81s on average for producing a WSI stitch, an order of magnitude faster than the Pure Graph (7512.38s). While global graph alignment without optical flow neighborhoods needs $O(n^2)$ pairwise comparisons to compute translations against all possible pairs, GloFlow only needs $O(n)$ with neighborhood information. The optical flow techniques in isolation are faster than GloFlow at 249.60s (LK) and 77.35s (FNM), but the stitches produced by GloFlow are observed to be more accurate (Figure \ref{fig:main}).

\begin{table}[t]
    \resizebox{\columnwidth}{!}{
\begin{tabular}{|crrr|}
\Xhline{2\arrayrulewidth}
Technique    & EPE   & Re-EPE  & Time (s)  \\ 
\Xhline{2\arrayrulewidth}
\textbf{Pairwise} & & & \\
LK & 12.52 & 1386.08 & 249.60 \\
FNM & \textbf{6.92}  & 745.80  & 77.35 \\
\hline
\textbf{Global} & & & \\
Pure Graph   &   N/A &  507.09   & 7512.38   \\
GloFlow (LK) (ours)  & N/A   & \textbf{50.99}   & 857.32 \\
GloFlow (FNM) (ours)   & N/A   & 51.14   & \textbf{684.81} \\ \hline
\end{tabular}
}
\caption{Comparison of image stitching methods.}
\label{tab:table1}
\end{table}

\section{Conclusion}
The purpose of this work was to develop a method to create an accurate whole slide image stitch from a video scan in a computationally tractable manner. We introduce GloFlow, a method that combines optical flow with graph alignment, and demonstrate superior performance to optical flow and graph based methods on a simulated dataset of pathology slide video scans.
We hope that this work will lower the barrier to digitization in pathology enabling access to digital pathology in low resource settings. 

\bibliography{gloflo}

\appendix

\section{Variation in Data Generation}\label{apd:first}

Variations in the data generation process are produced using four hyperparameters which govern the average displacements between patches and the approximate overlap between horizontal passes. The horizontal or vertical displacement between any two successive patches is defined by a magnitude and rotational angle centered along the direction of displacement (either right, left or down). The average magnitude of the displacement is defined as a factor of the patch width and is sampled uniformly between 14.62 and 20.48 (512/35 and 512/25 respectively). Zero centered Gaussian noise is added to this magnitude with the standard deviation defined as the given velocity divided by a noise factor where the noise factor is itself defined globally for a single data generation process and is sampled uniformly between 25 and 5. The angle of the displacement is sampled for  a Gaussian distribution centered at 0 with standard deviation sampled once uniformly between 1 and 15 degrees.  The overlap between successive rows is sampled uniformly between 0.2 to 0.4 to correspond to an approximate 20\% to 40\% overlap. For each data generation we save an array of translations between successive patches which serve as the ground truth.

\section{Re-EPE Equations}\label{apd:second}

\begin{equation}
    \begin{aligned}[b]
    L_{\text{RE-EPE}} = \frac 1N \sum_i L_{\text{EPE}}
    (P - P[i]\mathbf{1}, \\ T - T[i] \mathbf{1})
\end{aligned}
\end{equation}

\begin{equation}
    \begin{aligned}[b]
    L_{\text{EPE}}(P, T) = \sum_i
    \sqrt{(P[i] -  T[i])^2}
\end{aligned}
\end{equation}

where $P$ and $T$ are the predicted and ground truth locations vectors and the sum is over every node $i$.

\end{document}